# Coulomb effects in artificial molecules

Felipe Ramirez†

*Centro de Investigación Científica y de Educación Superior de Ensenada, Ensenada, B.C., México*

E. Cota

*Instituto de Física - UNAM, Laboratorio de Ensenada, Ensenada, B.C., México*

Sergio E. Ulloa

*Department of Physics and Astronomy and CMSS Program, Ohio University, Athens, OH 45701-2979*

We study the capacitance spectra of artificial molecules consisting of two and three coupled quantum dots from an extended Hubbard Hamiltonian model that takes into account quantum confinement, intra- and inter-dot Coulomb interaction and tunneling coupling between all single particle states in nearest neighbor dots. We find that, for weak coupling, the interdot Coulomb interaction dominates the formation of a collective molecular state. We also calculate the effects of correlations on the tunneling probability through the evaluation of the spectral weights, and corroborate the importance of selection rules for understanding experimental conductance spectra.

## 1. Introduction

An artificial atom or quantum dot is a fabricated nanostructure in which electrons have been confined in all three dimensions, leading to a discrete energy spectrum of the electronic states. The electrostatic charging energy $U = e^2/C$ (C is the total capacitance) is about an order of magnitude larger than the average level spacing $\Delta\epsilon$, leading to interesting new phenomena due to the interplay between energy and charge quantization [1, 2]. If the artificial atom is coupled to electron reservoirs, one can study transport properties through the quantum dot. At low bias voltage and low temperatures ($kT << \Delta\epsilon$), periodic oscillations of the conductance as a function of gate voltage have been observed [1] and explained in terms of the Coulomb blockade effect [3, 4]. In this linear regime, the difference in gate voltage between two successive conductance peaks corresponds to the difference in energy between the ground states for $N-1$ and $N$ electrons, and is called the addition energy. Transport spectroscopy through these systems has been carried out [5] by varying both the gate voltage and the bias voltage. In the non-linear regime, the bias voltage is finite and excited states are accessible, providing additional transport channels.

† Permanent address: Centro de Graduados e Investigación, Instituto Tecnológico de Tijuana, Apartado Postal 1166, Tijuana, B.C., México

    

In a seminal paper, Meir *et al.* [6] studied the temperature dependence of the conduction peaks by using an Anderson Hamiltonian with a Hubbard term for the intradot electron-electron interaction. Tunneling selection rules introduced by electronic correlations have been studied in the nonlinear regime by Weinmann *et al.* and by Pfannkuche and Ulloa [7], and the connection with the mechanism which appears to suppress many of the conductance peaks observed in experiments has been analyzed.

Recently, experiments have been done in arrays of two and three coupled quantum dots (artificial molecules), where the interdot coupling can be controlled [8], and the unfolding of each conductance peak into two and three peaks, respectively, is clearly observed as the coupling is increased. Transport experiments on artificial crystals show evidence for energy band formation [9]. The electron-electron interaction manifested in artificial atoms in the Coulomb blockade regime is expected to be equally important in arrays of quantum dots in the tunneling regime. Stafford and Das Sarma [10] studied the addition spectra of arrays of quantum dots to analyze the analog of the Mott metal-insulator transition and to explain experimental results [9] in 1D arrays of quantum dots. Other studies have been carried out using this approach, looking at the conductance in linear arrays of up to six quantum dots [11].

In this paper we study arrays of quantum dots using an extended Hubbard Hamiltonian that takes explicitly into account the interdot Coulomb repulsion and its effects on the addition spectrum. Also, we study the interplay between electron correlations and interactions through the evaluation of the overlap or spectral weights governing the tunneling probability through these systems.

## 2. Model

In the extended Hubbard Hamiltonian, besides the intradot charging energy $U$, we consider a more general tunneling matrix element $t_{\alpha\beta}$ between single particle states in nearest-neighbor dots and the interdot Coulomb interaction $V$, invariably present in experiments. The Hamiltonian can be written as

$$\hat{H} = \sum_{j,\alpha} \epsilon_{j,\alpha} \hat{C}^{\dagger}_{j,\alpha} \hat{C}_{j,\alpha} - \sum_{\alpha\beta,<i,j>} (t_{\alpha\beta} \hat{C}^{\dagger}_{i,\alpha} \hat{C}_{j,\beta} + h.c.)$$
$$+ \frac{1}{2} \sum_j U_j \hat{n}_j (\hat{n}_j - 1) + \sum_{<i,j>} V_{ij} \hat{n}_i \hat{n}_j, \qquad (1)$$

where $\hat{n}_j$ is the electron number operator at site $j$ and $<i,j>$ represents nearest neighbor dots. In this equation, $\epsilon_{j,\alpha}$ are the energy levels of the $j^{th}$ quantum dot which are assumed equally spaced with separation $\Delta_j$, $U_j$ is the intradot Coulomb repulsion for the $j^{th}$ quantum dot, $V_{ij}$ is the interdot repulsion between the $i^{th}$ and $j^{th}$ dots and $t_{\alpha\beta}$ is the tunneling matrix element between $\alpha$ and $\beta$ orbital states. Hereafter, energies will be expressed in terms of $U_1 = U$ and $V_{ij}$ is assumed constant and equal to $V$.

The tunneling matrix elements $t_{\alpha\beta}$ will be assumed to either be zero, or be given by $t_{\alpha\beta} = \gamma^2 t/((\Delta E)^2 + \gamma^2)$, where $\Delta E = \epsilon_{i,\alpha} - \epsilon_{i+1,\beta}$ is the difference in energy between levels in each dot involved in the tunneling event, $t$ is the maximum value of the tunneling matrix element and $\gamma$ is the width of the Lorentzian (set here equal to $1/t$), This form of $t_{\alpha\beta}$ simulates the expected decreasing coupling between levels that are not resonant. A specific microscopic model of the structure allows one to estimate the tunneling matrix elements and obtain this behavior. Details will be presented elsewhere.

In all cases, we have taken the temperature such that $kT = 0.04U$, and $U > t \sim \Delta_j >> kT$ for all $j$.

The current through the system takes into account the fact that an electron entering or leaving the system causes a transition between the $(N-1)$-electron state $|\alpha'>$ and the $N$-electron state $|\alpha>$. The corresponding tunneling rates $\Gamma_{\alpha\alpha'} = \Gamma_n S_{\alpha\alpha'}$ depend on the single electron tunneling rates $\Gamma_n$ and the spectral weights $S_{\alpha\alpha'}$ describing the correlations in the system. This quantity is given by [7]

$$S_{\alpha\alpha'} = \sum_n |<\alpha|\hat{C}_n^\dagger|\alpha'>|^2. \tag{2}$$

where $n$ labels single-electron dot states. For a system of uncorrelated electrons this overlap will be either one or zero between any two states; however, electron correlations result in overlaps much less than unity and consequently the tunneling probability is reduced considerably.

Here we present results for linear arrays of two and three quantum dots with up to five spin degenerate levels per dot. For the two atom molecule we consider two cases: (a) the symmetric case, where the level spacing and intradot Coulomb interaction are the same in each site, i.e., $\Delta_1 = \Delta_2 = \Delta$ and $U_1 = U_2 = U$; (b) the asymmetric case, where these parameters are different in each site. We assume a parabolic potential for each dot, so that these two parameters characterize the size of each atom. For the three atom molecule, we consider linear arrays in the following two ways: (a) all atoms are equal and (b) a larger atom in the center, with appropriate parameters for each case.

The procedure is to solve the extended Hubbard Hamiltonian in the particle number representation by direct diagonalization to obtain the eigenvalues and eigenvectors for the system with $N$ electrons. From the eigenvalues we can obtain the addition spectrum from

$$\frac{\partial <N>}{\partial \mu} = kT \frac{\partial^2 \ln \mathcal{Z}}{\partial \mu^2}, \tag{3}$$

where $\mathcal{Z}$ is the grand canonical partition function and the differential capacitance is obtained from

$$\frac{\partial Q}{\partial \mathcal{V}} = e^2 \frac{\partial <N>}{\partial \mu}. \tag{4}$$

The spectral weights can be calculated from the eigenstates of $N-1$ and $N$ electrons and equation (2).

## 3. Results

For two dots, in the symmetric case, with $t_{\alpha\beta} = 0$, $V = 0$ (Fig. 1(a)), we obtain peaks in the differential capacitance spectrum, characteristic of isolated dots, separated by the Coulomb blockade energy $\Delta + U$. The dotted line shows the behavior of $<N>$, the average number of electrons in the molecule where the degeneracy present in our system can be observed. Keeping the tunneling coupling equal to zero and increasing the interdot Coulomb repulsion $V$ we observe (Fig. 1(b)) that the Coulomb blockade is partially destroyed with an unfolding of the peaks, with separation equal to the value of $V$, in such a way that a doublet is obtained from each of the original peaks. By increasing the value of $V$ further, we observe the evolution of the spectrum from the "atomic" case of isolated quantum dots, to a collective state of the system where both the intra- and interdot Coulomb interactions, $U$ and $V$, are present, but it is $V$ which is responsible for breaking the degeneracy and producing these collective states.

In the asymmetric case, Fig. 1(c), we have a different situation where the peak at $\mu = 0$ characteristic of isolated dots remains as a consequence of the alignment of the lowest energy levels for both dots, but the other two peaks unfold due only to the difference in size between the dots. Again we observe the formation of split peaks as $V$ is increased (Fig. 1(d)).

When $t = 0.1U$ and $V = 0$, in the symmetric case, Fig. 2(a), the splitting of the peaks is proportional to the value of $t$. For weak coupling we see that the peaks have not completely unfolded; but, as $V$ is increased (Fig. 2(b)) again we observe the formation of split off features. Comparing with Fig. 1(b), we note that although both plots look similar, there is an important difference. Here there are strong electron correlations not present in the situation corresponding to Fig. 1(b). Therefore, we may say that capacitance spectra do not yield information on effects of correlations. As we will see below, we may obtain this information through the calculation of spectral weights. In the asymmetric case, for $V = 0$ (Fig. 2(c)) again we see a splitting of the peaks which now depends on both the tunneling coupling and the asymmetry, and as the interdot interaction $V$ is turned on, we observe the formation of a series of collective states of the system (Fig. 2(d)).

Figures 3(a) and (b) show the spectra for molecules with three identical dots and three orbitals, with interdot tunneling ($t = 0.1U$), for the cases $V = 0$ and $V = 0.3U$. Since we are restricted to interactions between nearest neighbors, we obtain results similar to the corresponding case for two dots, shown in Figs. 2(a) and 2(b). Considering an atom with a larger size at the center of the molecule (Figs. 3(c) and 3(d)), we see that the reduction in symmetry improves the formation of collective molecular states. As the number of dots increases we observe how each peak in the capacitance spectrum evolves into a "miniband" consisting of as many peaks as dots in the artificial molecule. One notes gaps between successive minibands which in general decrease as the coupling ($t$ or $V$) is increased. The capacitance spectrum exhibits a collective property or state which is a signature of the artificial molecule in a similar sense as the energy band structure defines a system in solid state physics. The capacitance peaks correspond to fluctuations in $N$, the number of electrons in the system, determined by changes in chemical potential $\mu$ as the gate voltage is varied in transport spectroscopy or capacitance experiments, and obey selection rules which depend on the electron correlations in the system.

We see from these results that, for weak tunneling, the effect of interdot Coulomb interaction is very important in the formation of collective states in coupled quantum dot arrays. The general behavior we observe for the evolution of the differential capacitance peaks as a function of coupling, with interaction effects present, are in agreement with experimental results [8], regarding the position of the conductance peaks.

In Figure 4, we show results for the spectral weights $S_{\alpha\alpha'}$ as a function of the energy difference $\Delta E$ between the states involved in the transition, corresponding to the case where the number of electrons in the system goes from $N = 2$ to $N = 3$ keeping the interdot Coulomb interaction $V = 0.3U$ fixed. For $t_{\alpha\beta} = 0$, Figure 4(a) shows that for two symmetric quantum dots with three orbitals, the electrons are totally uncorrelated. This can be understood since in our model Hamiltonian the interaction terms are diagonal. When $t = 0.1U$, we see from Fig. 4(b) for the same system, an overall reduction in the values of the spectral weights due to the presence of correlations, since in this case a state of the system is made up from a large number of single particle states with very different occupation probabilities. Figure 4(c) shows the spectral weights corresponding to the case in Fig. 3(c), and Fig. 4(d) for an asymmetric double dot system with five orbitals. We can see that a large number of overlaps with small values appears, implying that in this system most of the channels will have a small contribution in transport experiments. These are the effects of electron correlations caused by tunneling coupling which cannot be appreciated in the corresponding addition spectra (Fig. 1(b), 2(b)). If we calculate the average value of the overlap for these two cases we obtain 0.083 and 0.067, respectively

## 4. Conclusions

We investigate artificial molecules consisting of two and three quantum dots coupled in series. The

*Superlattices and Microstructures, Vol. 20, No. 1, 1996* 5dots can be either equal or with different sizes. We have applied an extended Hubbard Hamiltonian to calculate the capacitance spectra and spectral weights for these systems. We find that the peak positions depend on both inter- and intra-dot interactions and observe how characteristic peaks of isolated dots develop into minibands and a series of molecular "collective states" for $N$ electrons is formed. We observe that the interdot Coulomb repulsion must be considered in the weak tunneling regime for an appropriate interpretation of peak splitting in conductance spectra experiments. We analyze the spectral weights for these artificial molecules to emphasize the interplay between interactions and electron correlations and the relation with the tunnel probabilities through these systems. Our results are in agreement with known selections rules for quantum dots.

*Acknowledgments.* This work was supported in part by CONACYT project 400363-5-0078PE, DGAPA-UNAM project IN-100895 and US–DOE grant No. DE–FG02–91ER45334. FR acknowledges a scholarship from CONACYT, México.# References

[1] J. H. F. Scott-Thomas, S. B. Field, M. A. Kastner, H. I. Smith and D. A. Antoniadis, Phys. Rev. Lett. **62**, 583 (1989); U. Meirav, M. A. Kastner and S. J. Wind, Phys. Rev. Lett. **65**, 771 (1990).
[2] H. van Houten, C. W. J. Beenakker, and A. M. Staring, in *Single Charge Tunneling*, H. Grabert and M. H. Devoret, eds., NATO ASI Ser. B **294** (Plenum Press, NY, 1991); M. A. Kastner, Physics Today **46**, 24 (1993).
[3] D. V. Averin and K. K. Likharev, in *Mesoscopic Phenomena in Solids*, B. L. Altshuler, P. A. Lee, R. A. Webb, eds. (Elsevier, Amsterdam, 1991).
[4] H. van Houten and C. W. J. Beenakker, Phys. Rev. Lett. **63**, 1893 (1989).
[5] P. L. McEuen, E. B. Foxman, J. Kinaret, U. Meirav, M. A. Kastner, N. S. Wingreen and S. J. Wind, Phys. Rev. B **45**, 11419 (1992); A. T. Johnson, L. P. Kouwenhoven, W. de Jong, N. C. van der Vaart, C.J.P.M. Harmans and C. T. Foxon, Phys. Rev. Lett. **69**, 1592 (1992); J. Weis, R. J. Haug, K. v. Klitzing and K. Ploog, Phys. Rev. Lett. **71**, 4019 (1993).
[6] Y. Meir, N.S. Wingreen and P.A. Lee, Phys. Rev. Lett. **66**, 3048 (1991).
[7] D. Weinmann, W. Häusler and B. Kramer, Phys. Rev. Lett. **74**, 984 (1995); D. Pfannkuche and S.E. Ulloa, Phys. Rev. Lett. **74**, 1194 (1995)
[8] F.R. Waugh, M.J. Berry, D.J. Mar, R.M. Westervelt, K.L. Campman and A.C. Gossard, Phys. Rev. Lett. **75**, 705 (1995); R.H. Blick, R.J. Haug, J. Weis, D. Pfannkuche, K. v. Klitzing and K. Eberl, Phys. Rev. B **53**, 7899 (1996)
[9] L. P. Kouwenhoven, F. W. J. Hekking, B. J. van Wees, C. J. P. M. Harmans, C. E. Timmering and C. T. Foxon, Phys. Rev. Lett. **65**, 361 (1990); R.J. Haug, J.M. Hong and K.Y. Lee, Surf. Sci. **263**, 415 (1992)
[10] C.A. Stafford and S. Das Sarma, Phys. Rev. Lett. **72**, 3590 (1994)
[11] G. Chen, G. Klimeck, S. Datta, G. Chen and W. A. Goddard III, Phys. Rev. B **50**, 8035 (1994)**Fig. 1.** Differential capacitance (in arbitrary units) vs chemical potential for a molecule with two quantum dots and three single particle states in each dot. No interdot tunneling and varying interdot Coulomb repulsion $V$. For a symmetric molecule with $U = 1$ and $\Delta = 0.3$ in each dot, (a) $V = 0$, (b) $V = 0.3$. For an asymmetric molecule with $U_1 = 1$, $\Delta_1 = 0.3$; $U_2 = 0.8$, $\Delta_2 = 0.2$, (a) $V = 0$, (b) $V = 0.3$

**Fig. 2.** Same as Fig. 1, with interdot tunneling present ($t = 0.1$).

**Fig. 3.** Differential capacitance vs chemical potential for a molecule with three quantum dots and three single particle states in each dot. Interdot tunneling is present ($t = 0.1$) and varying interdot Coulomb repulsion $V$. Identical quantum dots, $U = 1$, $\Delta = 0.3$ in each dot, (a) $V = 0$, (b) $V = 0.3$. A molecule with a larger atom in the center and parameters $U_1 = U_3 = 1$, $\Delta_1 = \Delta_3 = 0.3$; $U_2 = 0.8$, $\Delta_2 = 0.2$, (c) $V = 0$, (d) $V = 0.3$

**Fig. 4.** Spectral weights as a function of transition energy $\Delta E$ between states with $N = 2$ and $N = 3$ electrons. Interdot Coulomb repulsion is $V = 0.3$ ($U = 1$). (a) Symmetric molecule with two dots and three orbitals with the tunneling interaction $t_{\alpha\beta} = 0$, and (b) $t = 0.1$, to contrast the role of electron correlations and its effect on spectral weights. (c) A system of three dots and three orbitals with the larger dot in the center and $t = 0.1$. (d) Asymmetric molecule of two dots and five single particles energy levels in each dot and $t = 0.1$.

Fig. 1
Coulomb effects in artificial molecules
F. Ramirez, E. Cota and S.E. Ulloa

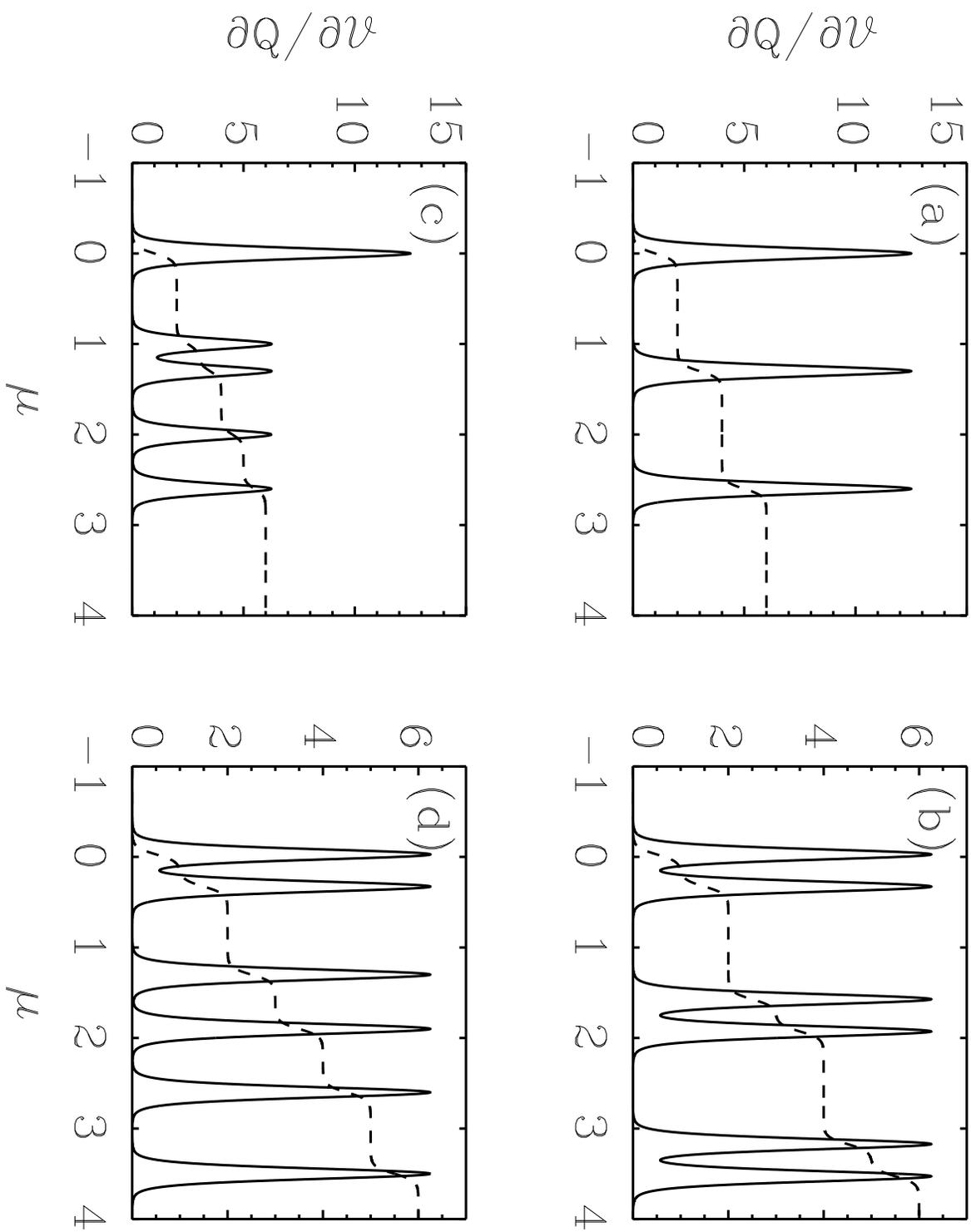

Fig. 2
Coulomb effects in artificial molecules
F. Ramirez, E. Cota and S.E. Ulloa

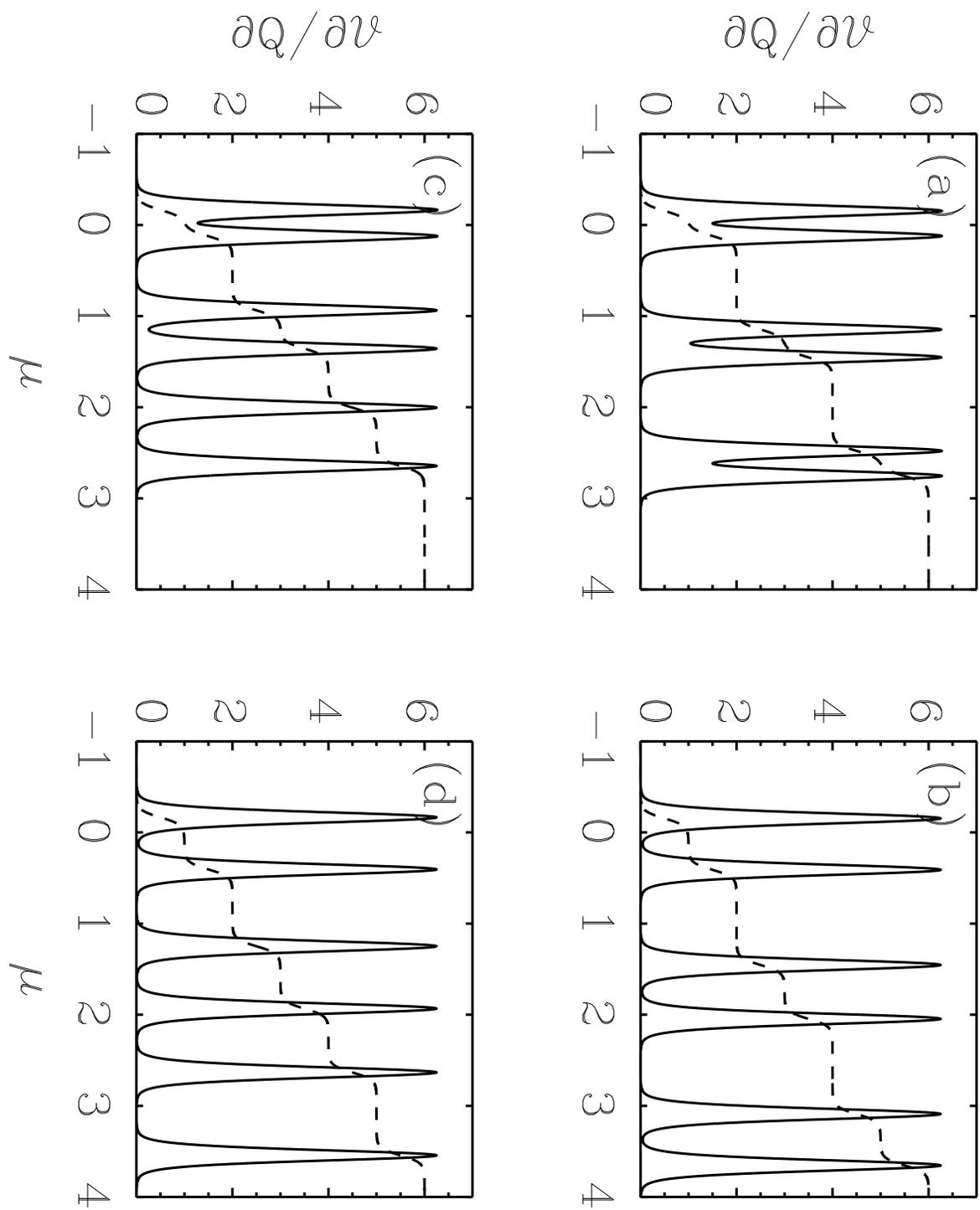

Fig. 3
Coulomb effects in artificial molecules
F. Ramirez, E. Cota and S.E. Ulloa

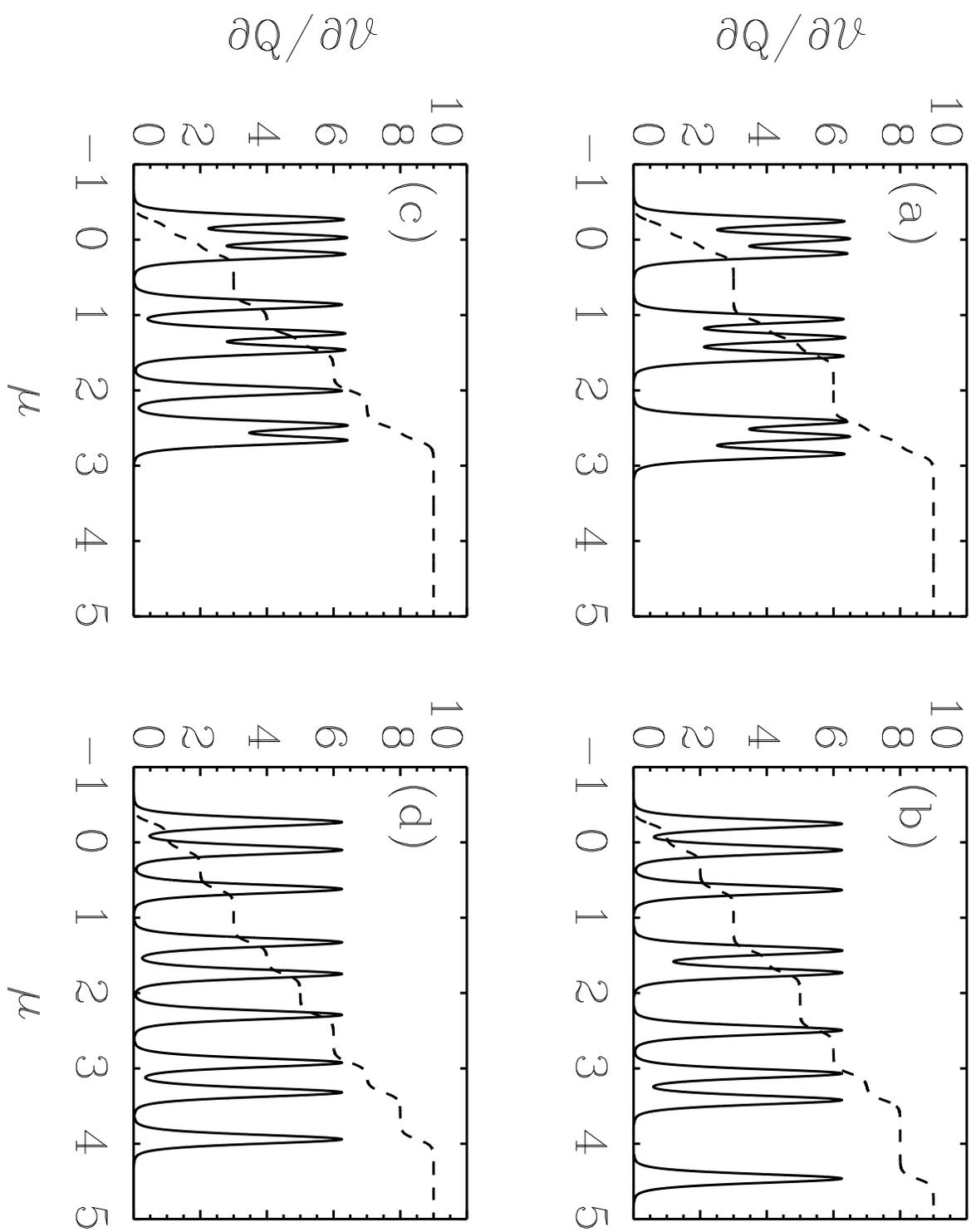

Fig. 4
Coulomb effects in artificial molecules
F. Ramirez, E. Cota and S.E. Ulloa

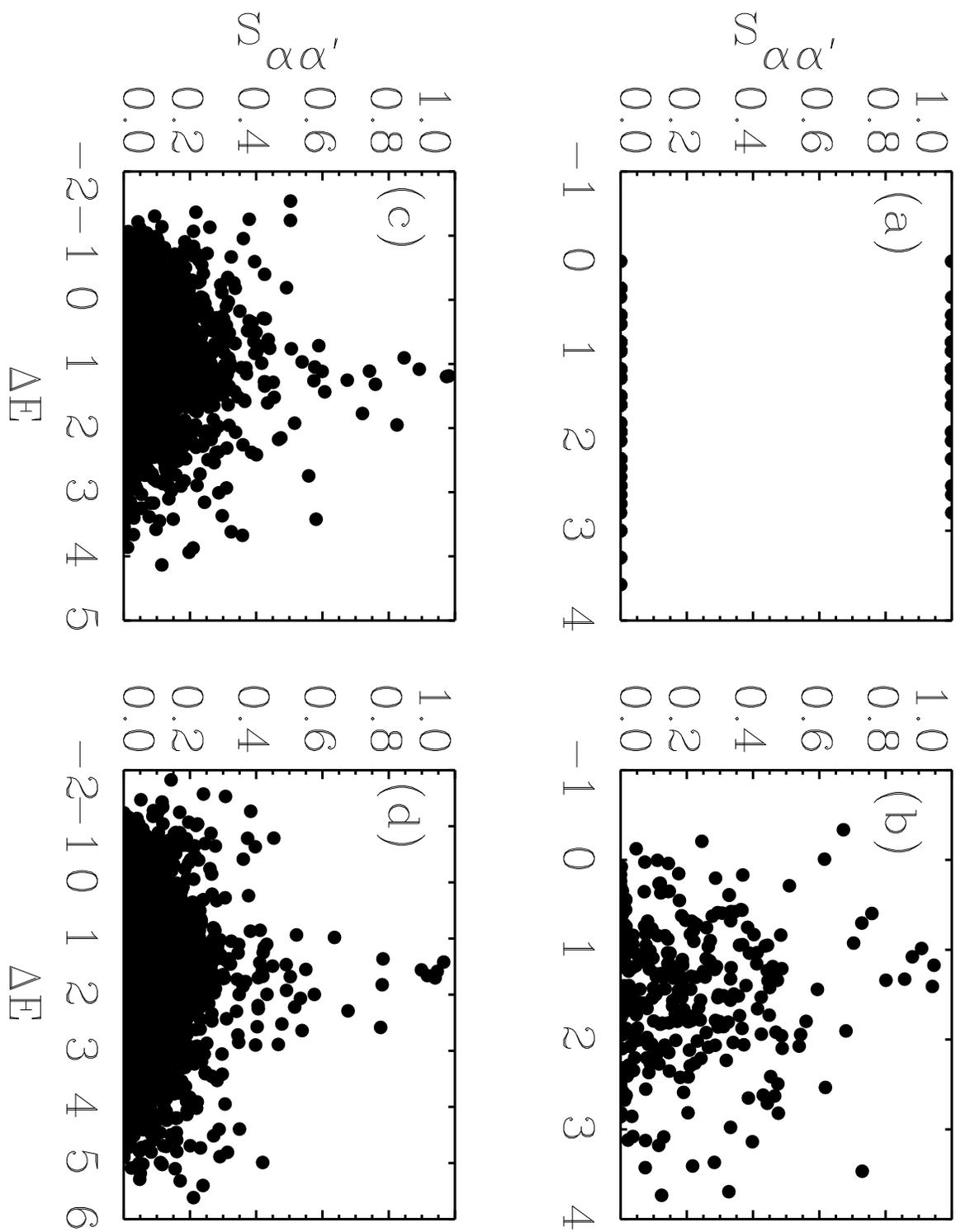